# Robust skyrmion mediated reversal of ferromagnetic nanodots of 20 nm lateral dimension with high M$_s$ and moderate DMI

Md Mahadi Rajib[1], Walid Al Misba[1], Dhritiman Bhattacharya[1] and Jayasimha Atulasimha[1,2]

[1]*Department of Mechanical and Nuclear Engineering, Virginia Commonwealth University, Richmond, VA 23284, USA*

[2]*Department of Electrical and Computer Engineering, Virginia Commonwealth University, Richmond, VA 23284, USA*

\* Corresponding author: jatulasimha@vcu.edu

**ABSTRACT**

Implementation of skyrmion based energy efficient and high-density data storage devices requires aggressive scaling of skyrmion size. Ferrimagnetic materials are considered to be a suitable platform for this purpose due to their low saturation magnetization (i.e. smaller stray field). However, we show by performing rigorous micromagnetic simulation that such scaling of skyrmion size by lowering saturation magnetization while applicable in infinite films or where the skyrmion size is very small compared to the film's lateral dimension, does not hold in confined geometries. We also found in confined geometries, where skyrmion occupies the whole volume of a nanodot, high saturation magnetization helps form stable skyrmions. Specifically, such skyrmions can be formed in 20 nm lateral dimension nanodots with high saturation magnetization (1.6-1.71 MA/m) and moderate DMI (3 mJ/m$^2$). This result could stimulate experiments on implementation of highly dense skyrmion devices. In particular, we show that Voltage Controlled Magnetic Anisotropy (VCMA) based switching mediated by an intermediate skyrmion state can be achieved in the soft layer of a ferromagnetic p-MTJ of lateral dimensions 20 nm with sub 1fJ/bit energy in the presence of room temperature thermal noise with reasonable DMI ~3 mJ/m$^2$.

**Introduction**

Skyrmions are particle like localized spin structures which can potentially overcome pinning with much smaller currents compared to domain walls (DW) [1,2,3,4,5,6,7]. This characteristic provides a pathway to implement energy efficient racetrack devices [7,8,9,10]. Skyrmions confined in the free layer of a Magnetic Tunnel Junction (MTJ) switched by electrical fields can also function as a memory device [11, 12]. For example, skyrmions can assist voltage controlled magnetic anisotropy (VCMA) switching of perpendicular ferromagnets from up/down state to down/up state acting as an intermediate state [13,14,15]. This skyrmion mediated VCMA reversal has been shown to be robust and energy efficient while not requiring a bias magnetic field. This could lead to robust, scalable and energy efficient (<1fJ/bit) VCMA switched MTJs. However, aggressive scaling of skyrmion size, both in case of racetrack and MTJ devices, is required to make them competitive with existing STT-MRAM devices in terms of density and energy efficiency.

Multilayer thin ferromagnetic film stacks have been optimized to reduce skyrmion size down to ~50 nm starting from several micrometer lateral dimensions at room temperature [16,17,18]. Unfortunately, large stray fields originating from ferromagnets impede further scaling of skyrmions in thin films [19]. In case of confined structures, it has been previously shown that creation, annihilation and dynamics of skyrmions is drastically different from thin films due to the influence of geometric boundary on skyrmion stability and dynamics [6]. Though there have been few studies on skyrmions in a confined structure [20,21,22,23], no prior work studies the requirement on both saturation magnetization ($M_s$) and Dzyaloshinskii-Moriya Interaction (DMI) as the lateral dimension is downscaled to very small sizes ~20nm. We previously showed ferromagnetic skyrmions cannot be formed in a 20 nm nanodot with experimentally observed DMI (~3.3mJ/m$^2$ [24]) and require extremely large DMI (~13mJ/m$^2$) [15]. Hence, compensated ferrimagnets have recently emerged as an alternative to ferromagnets for hosting smaller skyrmions due to their small stray fields [25,26]. In ferrimagnetic thin film of $Co_{44}Gd_{56}$, ~10 nm skyrmion has been observed without any out-of-plane bias field with lifetime on the order of several minutes [25]. However, in this work we show that high saturation magnetization favors formation of skyrmions in confined structures of small lateral dimensions (e.g. nanodots ~20 nm) as opposed to the case of ferrimagnetic thin films where weak stray fields due to low net saturation magnetization helps form smaller skyrmions. We also show that thermally robust skyrmion mediated switching can be attained in a 20 nm nanodot with high saturation magnetization and DMI values ~3mJ/m$^2$ that have been experimentally observed.

## Results

### Formation of skyrmion in smaller nanodots requires higher $M_s$ and DMI (simulations at 0K)

We simulated the magnetization dynamics of circular nanodots of 100 nm, 30 nm and 20 nm lateral dimensions using the micromagnetic simulation software Mumax3 [27]. Details of the simulation and parameters used are discussed in the methods section. In all three cases, we initiated the simulations from a purely ferromagnetic state (all the spins pointing in the +z direction) and observed the final equilibrium states after 5ns which was found to be sufficient for reaching a stable state. This was confirmed by studying one equilibrium skyrmion state ($M_s$=1.7MA/m, DMI=3.3 mJ/m$^2$) for 1 µs without any change in the magnetization beyond 5 ns.

To find minimum DMI value for which skyrmions are formed as the lateral dimension is reduced from 100 nm to 20 nm, $M_s$ and DMI were varied under a constant effective PMA energy. The saturation magnetization was varied from 1.0 MA/m to 1.8 MA/m in all three cases. We varied the perpendicular anisotropy ($K_{u1}$) as a function of saturation magnetization in such a way that effective anisotropy $K_{eff}V = \left(K_{u1} - \frac{1}{2}\mu_0 M_s^2\right)V$ always remains a constant for all nanodots. We noticed that starting from a purely ferromagnetic state, three different states were stabilized, namely skyrmion (SK), multidomain (MD) and quasi-ferromagnetic (QFM) states (Fig. 1). Total energy (Fig. 2a, b, c) of each of the state is studied as a function of $M_s$ and DMI and their corresponding topological charge ($\frac{1}{4\pi} \int \vec{m}.\left(\frac{\delta \vec{m}}{\delta x}\right) \times \left(\frac{\delta \vec{m}}{\delta y}\right) dx\, dy$) (Fig. 2d, e, f) is also observed for the verification of skyrmion state. We discuss these in details in the next sections.

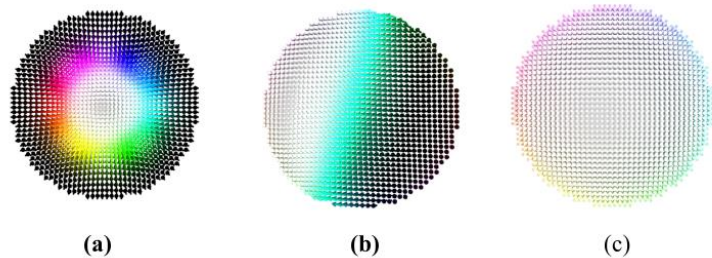

**Figure 1.** Three stable states: (a) Skyrmion, (b) Multidomain and (c) Quasi-ferromagnet

First, we considered 100 nm nanodots and studied the saturation magnetization values for which the least amount of DMI is needed to form a skyrmion. To find this value of DMI, the DMI was varied until a skyrmion state emerged as an equilibrium state for a $M_s$ range of 1.0-1.8 MA/m. Figure 2a shows that total energy decreases with increasing $M_s$ without DMI (blue line) and this trend is qualitatively followed for DMI values up to 0.9 mJ/m$^2$ (green line). For 0.0 mJ/m$^2$ to 0.9 mJ/m$^2$ DMI, all of the states are single domain ferro/quasi-ferromagnetic states. For DMI value of 1.0 mJ/m$^2$, we observe that skyrmion is formed at 1.0 MA/m and from Fig. 2d we see that the state has topological charge ~1 which confirms this state is a skyrmion. Total energy decreases for saturation magnetization of 1.01 MA/m and 1.02 MA/m and these equilibrium states are also skyrmions as can be seen from figure 2d. When $M_s$ is increased to 1.03 MA/m, while the total energy decreases further, the state is a multidomain state instead of a skyrmion. For 1.04 MA/m $M_s$, the total energy increases sharply and the equilibrium state is quasi-ferromagnetic. Beyond 1.04 MA/m, the total energy decreases with increasing $M_s$ and all of the equilibrium states are quasi-ferromagnetic states. Therefore, 1.0 mJ/m$^2$ is the minimum DMI for 100 nm nanodot i. e. below 1.0 mJ/m$^2$ DMI, no skyrmion can be formed for saturation magnetization ranging from 1.0 MA/m to 1.8 MA/m with the above-mentioned effective PMA energy and parameters listed in table 1 (method section). However, when DMI is 1.0 mJ/m$^2$, the range of saturation magnetization for which skyrmion can be formed is 1.00-1.02MA/m.

We next reduced the lateral dimension to 30 nm and explored the minimum DMI required to form skyrmions in the 1.0-1.8 MA/m $M_s$ range at the effective PMA energy equal to that of 100nm. From Fig. 2b we can see that total energy decreases with increasing $M_s$ without DMI (blue line), similar to the 100nm case. When DMI is increased to 2.1 mJ/m$^2$, although some multidomain states appeared in the $M_s$ range of 1.21-1.26 MA/m, no skyrmions were observed. At DMI value of 2.2 mJ/m$^2$, the total energy increases with $M_s$, resulting in quasi-ferromagnetic states up to $M_s$=1.24MA/m. At 1.24MA/m, DMI dominated state (multidomain state or skyrmion state) appears and it prevails in the $M_s$ range of 1.25-1.38 MA/m. We notice from Fig. 2b that the multidomain state at 1.25MA/m has much lower energy than quasi-ferromagnetic states and further increase in $M_s$ causes a sharp increase in total energy. Figure 2e shows that the states for 1.26-1.36MA/m have topological charge ~1 indicating it is a skyrmion state. Total energy again decreases at 1.37 MA/m and multidomain state forms. We obtain another multidomain state at 1.38 MA/m but after that total energy again increases and we notice that the states beyond 1.38MA/m are quasi-ferromagnetic states. Thus, we see from Fig. 2b, e that in the 30 nm nanodot, skyrmions are formed in the $M_s$ range of 1.26-1.36 MA/m. This is higher than the $M_s$ range required for formation of skyrmion in 100 nm nanodot at minimum DMI, and the requirement of minimum DMI increased from 1.0 mJ/m$^2$ (for 100nm) to 2.2 mJ/m$^2$ with the reduction in lateral dimension to 30 nm.

Finally, we explored the minimum DMI requirement for 20 nm. We observed that a higher DMI (3.3mJ/m$^2$) is needed to form a skyrmion in 20 nm nanodot than 30 nm and the $M_s$ values for which the skyrmion form shift to the higher end of the $M_s$ range. From Fig. 2f we can see that skyrmions are formed in the $M_s$ range of 1.6-1.71MA/m indicated by a topological charge ~1.

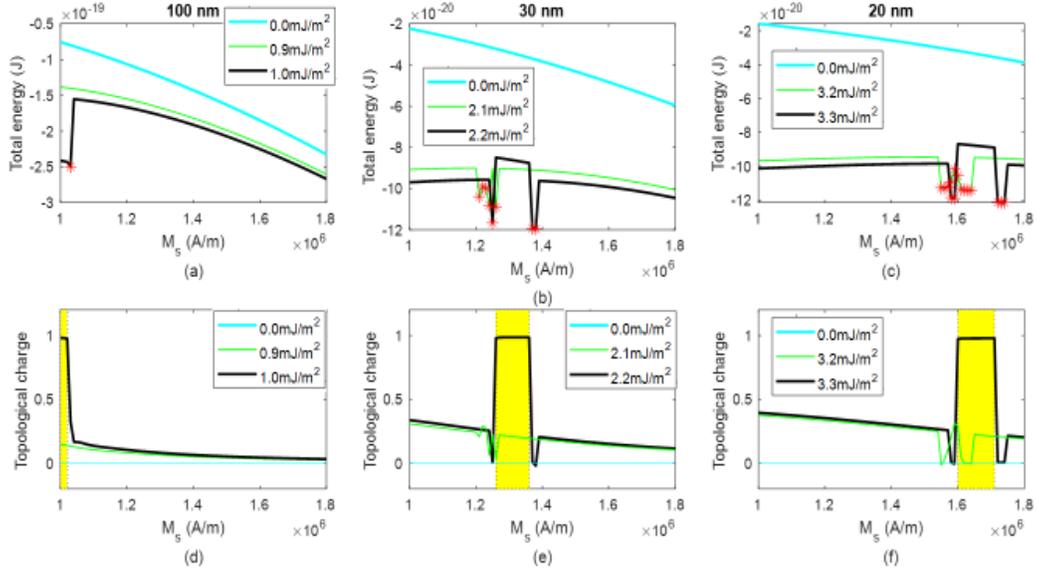

**Figure 2.** Total energy vs. saturation magnetization and corresponding topological charge at different DMI values for (a, d) 100 nm, (b, e) 30 nm and (c, f) 20 nm nanodots of 0.6 nm thickness (Blue line indicates the no DMI cases, asterisk sign represents multidomain state and area highlighted with yellow color indicates a skyrmion state).

We notice that in smaller nanomagnets (30nm and below), stabilized skyrmions are at a higher energy level compared to the quasi-ferromagnetic state and the multidomain state. To explore the reason behind this, we first studied the static energy of the three possible final states, quasi-FM, skyrmions and multidomain in 20 nm lateral dimension by varying the PMA and $M_s$ at a constant 3.3 mJ/m² DMI. As can be seen in Fig. 3, the three different states are at different energy levels where $E_{MD} < E_{QFM} < E_{SK}$. Although the multidomain state has the lowest energy for all the $M_s$ values studied, when starting from a purely ferromagnetic state, it is not always the equilibrium state. To elucidate this, we next studied the temporal evolution of energy during stabilization of these different equilibrium states starting from a purely ferromagnetic state.

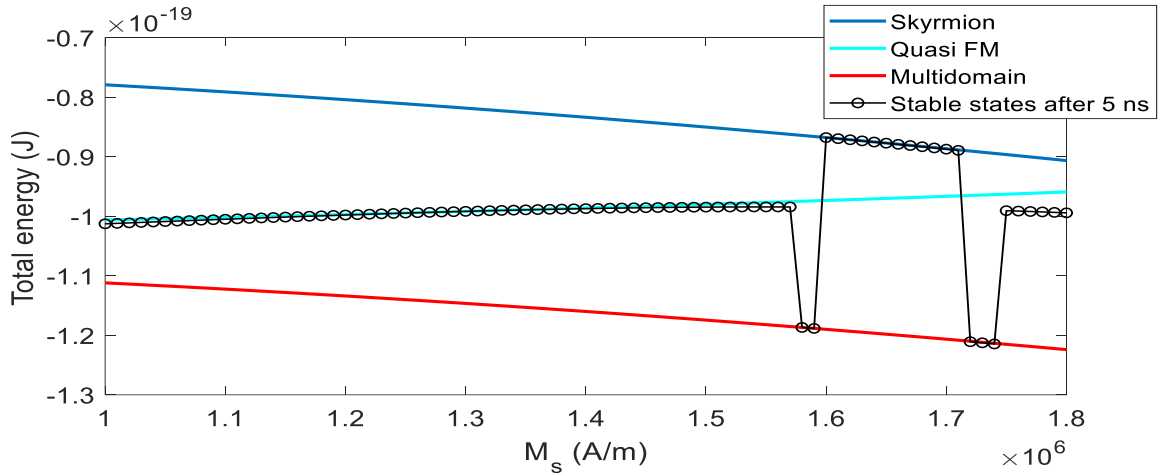

**Figure 3.** Comparison of the total energy of stable states after 5ns and the static energy of the skyrmion, quasi-FM and multidomain states at different $M_s$ values.

We considered three $M_s$ values for 20 nm nanodot, close to each other, particularly 1.57 MA/m, 1.58 MA/m and 1.60 MA/m and observed the dynamic evolution of the magnetic states and their corresponding energy as shown in Fig. 4. A skyrmionic state is formed in all of the three cases after ~50 ps of starting from a ferromagnetic state. After creation of the skyrmion state, breathing mode is excited. Skyrmion with 1.57 MA/m and 1.58 MA/m has the strongest breathing amplitude. Thus, during outbreathing, these skyrmions hit the boundary and are annihilated, creating a QFM state. While the QFM created for the 1.57 MA/m is stabilized, the QFM state created for 1.58MA/m has larger number of tilted spins at one side of the boundary which gyrotropically rotates and a multidomain state is ultimately formed. For $M_s$ value of 1.60 MA/m, the breathing mode is not strong enough. Therefore, the skyrmion does not annihilate at the boundary and stabilizes after few oscillations. Thus, these different states are all stable states and separated by finite energy barriers from each other. Specifically, although the skyrmions state has a total energy higher than the multidomain or QFM state, it is a local minimum with an energy barrier that allows it to remain a skyrmion, once the system enters the skyrmion state through a dynamic process. The final stabilized state is dependent on the strength of the breathing and the gyrotropic mode, which in turn depend on the system's magnetic parameters. We also note that as it is a dynamic process, different damping values can alter the range at which these different states are stabilized. However, the qualitative trends would be similar. Lastly, with thermal noise, one should expect a probabilistic distribution of these states at different $M_s$ values which could affect the switching behavior (as we investigate later).

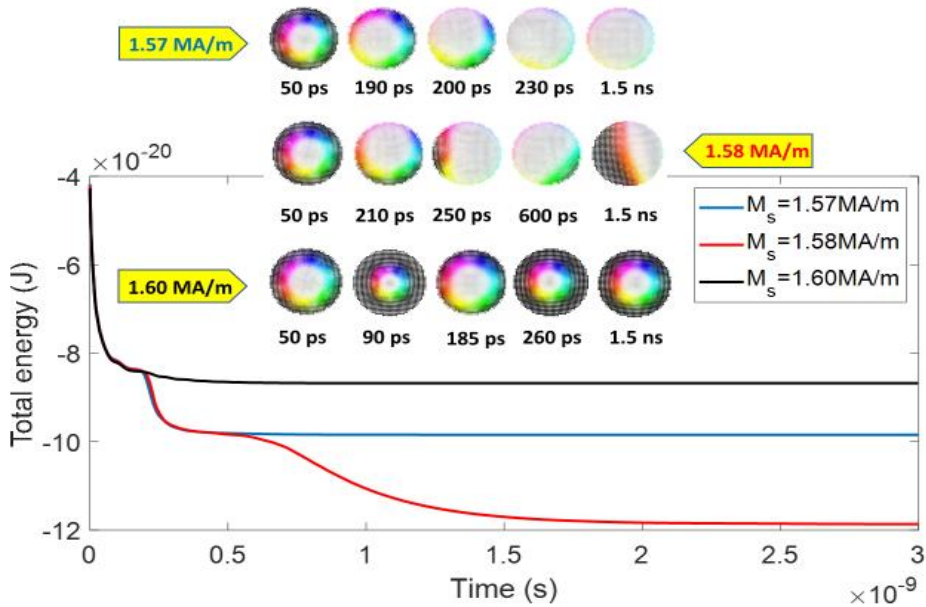

**Figure 4.** Evolution of total energy for three different $M_s$ values and their corresponding magnetic states at different times.

The requirement of higher DMI for decreased dimension of nanodots can be explained as follows. A higher DMI strength is required for completing the 360-degree rotation along a smaller diameter. Furthermore, when DMI is kept at the minimum required, high $M_s$ is needed as it helps the formation of stray field dominated skyrmions in the confined nanostructure. Figure 5 shows that for 20 nm nanodots and 3.3 mJ/m$^2$ DMI, minimum $M_s$ at which a skyrmion is formed is 1.60 MA/m whereas minimum $M_s$ for the formation of skyrmion at 4.0 mJ/m$^2$ DMI is only 1.34 MA/m. In summary, high $M_s$ is not needed if we could employ high DMI and therefore form DMI mediated skyrmions. On the other hand, using low DMI requires a high $M_s$ so that the stray field can aid the DMI in formation and stabilization of the skyrmions.

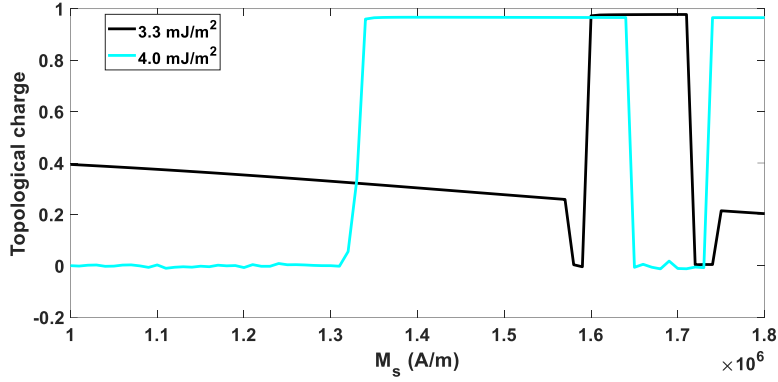

**Figure 5.** Topological charge vs. saturation magnetization of 20 nm nanodot for 3.3 mJ/m$^2$ and 4.0 mJ/m$^2$ DMI.

## Skyrmion-mediated voltage controlled switching of ~20nm nanodot (with inclusion of room temperature thermal noise)

In this section, we study the switching probability of 20 nm lateral dimension ferromagnetic nanodot by employing VCMA induced skyrmion mediated switching. In such switching, an intermediate skyrmion is created starting from a ferromagnetic state by lowering PMA which is subsequently annihilated by restoring the PMA to achieve switching from ferromagnetic up/down to down/up state (Fig. 6a). We have already discussed the formation and stability of a skyrmion in a confined ~20 nm nanodot with high $M_s$, experimentally observed DMI and low PMA in the previous sections which guides this investigation of skyrmion mediated switching of ferromagnets. Note that the $K_{eff}V$ under VCMA induced lowering of PMA, when the skyrmion is formed, does not determine the stability of the "bit". This is determined by the energy barrier when no VCMA is applied and PMA is the highest. The energy barrier of the ferromagnetic state is calculated by considering both the reduction due to demagnetization and DMI [28] and is expressed by $K_{eff\_MAX}V = (K_{u1\_MAX} - \frac{1}{2}\mu_0 M_s^2 - \frac{D^2 \pi^2}{16 A_{ex}})$ where $K_{u1\_MAX}$ is the perpendicular magnetic anisotropy constant for the ferromagnetic state, D is the DMI parameter and $A_{ex}$ is the exchange stiffness. Thus, the stability of the perpendicular ferromagnetic state is $K_{eff\_MAX}V/k_BT\sim 258$. The stability of the intermediate skyrmion state that is formed when the PMA is lowered by application of VCMA is needed for a short lifetime (100s of picosecond) to ensure robust switching of the ferromagnetic state through this intermediate skyrmion state [13].

For studying the switching probability, we considered $M_s$=1.7 MA/m, $A_{ex}$=6 pJ/m, Gilbert damping =0.01, and discretized the nanodot into 32×32×1 cells with each cell having volume of 0.625nm×0.625nm×1nm and initialized it in the ferromagnetic "up" state. We let the system relax for 500 ps to get the equilibrium state before trying to switch the magnetization. Due to the large PMA, this state is very close to the ferromagnet state with very small canting of peripheral spins due to stray field. We then applied VCMA to reduce the PMA from 6.33 MJ/m$^3$ to 1.85 MJ/m$^3$ within 0.5 ps. A skyrmion was formed ~40 ps after the reduction of PMA and had a lifetime of ~60 ps. We then restored the PMA by withdrawing the voltage pulse at different points (times in formation and dynamics of the skyrmion state) and observed switching. In this switching event, sub 1 fJ energy is dissipated on application of a voltage pulse of 2.0 V with 2240 fJ/Vm VCMA coefficient for a 1-nm-thick free layer and a 1-nm-thick MgO layer with relative permittivity ~7. We note that a high VCMA coefficient is needed as we chose a high initial PMA to ensure

$K_{eff\_MAX}V/k_BT$~ 258. If we go with a smaller $K_{eff\_MAX}V/k_BT$~ 40, which suffices for Random Access Memory (RAM), a VCMA coefficient ~800fJ/Vm [29] would be sufficient. Fig. 6b shows the switching percentage vs. pulse width and we notice that ~98% switching can be attained in the pulse width range of 51ps-61ps. With better optimization of the material parameters higher switching percentage can be attained at high $M_s$ and experimentally observed DMI with sub fJ energy for each switching event.

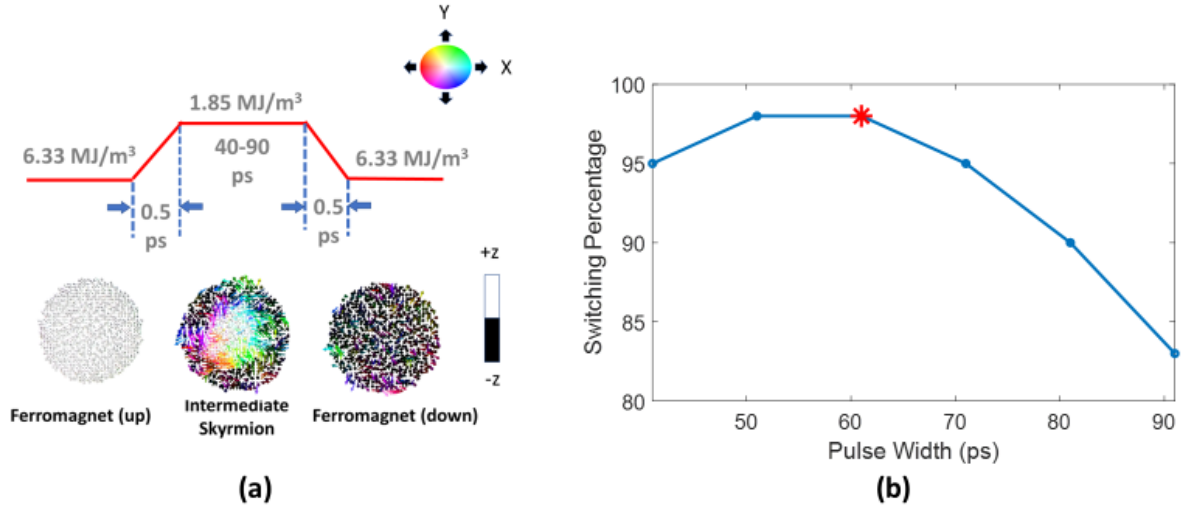

**Figure 6.** (a) Voltage pulse in terms of PMA and switching of 20 nm nanodot in the presence of thermal perturbation and (b) Switching percentage vs. pulse width (all of the points show the switching percentage for 100 simulation cases except the point marked with asterisk. The star mark corresponds to 61 ps pulse width where the simulations were run for 1000 times and 19 failures indicate ~98% switching).

**Discussion**

In summary, higher DMI can result in formation of skyrmions in smaller nanodots at low $M_s$ but such high values of DMI have not yet been experimentally observed. To create skyrmions in confined structures ~20 nm lateral dimensions with experimentally observed DMI ~3 mJ/m$^2$, one needs large demagnetization (stray field) from materials that can be achieved with a high $M_s$. We also showed that using a material with high saturation magnetization can help achieve thermally robust and extremely energy efficient switching in 20 nm ferromagnetic nanodot with experimentally demonstrated DMI. Thus, use of materials with high $M_s$ can provide a pathway for aggressive scaling of ferromagnetic skyrmion mediated VCMA switching of p-MTJs with lateral dimensions ~20 nm and beyond.

**Method**

The magnetization dynamics of circular nanodots was simulated by using the micromagnetic simulation software Mumax3 [27] for 100 nm, 30 nm and 20 nm lateral dimensions with a constant cell size of 0.5×0.5×0.6 nm$^3$. In the Mumax3 framework, the magnetization dynamics is simulated by solving the Landau-Lifshitz-Gilbert (LLG) equation:

$$\frac{\partial \vec{m}}{\delta t} = \left(\frac{-\gamma}{1+\alpha^2}\right) [\vec{m} \times \vec{B}_{eff} + \alpha\{\vec{m} \times (\vec{m} \times \vec{B}_{eff})\}] \tag{1}$$

where $\alpha$ is the Gilbert damping coefficient and $\gamma$ is the gyromagnetic ratio (rad/Ts). $\vec{m}$ is the normalized magnetization vector ($\vec{M}/M_s$), $M_s$ is the saturation magnetization and $\vec{B}_{eff}$ is the effective magnetic field having the following components [27]:

$$\vec{B}_{eff} = \vec{B}_{demag} + \vec{B}_{exchange} + \vec{B}_{dm} + \vec{B}_{anis} + \vec{B}_{thermal} \tag{2}$$

Here, $\vec{B}_{demag}$ is the effective field due to demagnetization energy and $\vec{B}_{exchange}$ is the Heisenberg exchange interaction.

$\vec{B}_{dm}$ yields the Dzyaloshinskii-Moriya interaction:

$$\vec{B}_{dm} = \frac{2D}{M_s}\left(\frac{\delta m_z}{\partial x}, \frac{\partial m_z}{\delta y}, -\frac{\partial m_x}{\delta x} - \frac{\partial m_y}{\delta y}\right) \tag{3}$$

Where $m_x$, $m_y$ and $m_z$ are the normalized magnetization components along the three cartesian co-ordinates and D represents the DMI constant (mJ/m$^2$).

$\vec{B}_{anis}$, the effective field due to perpendicular anisotropy, is given by the following equation:

$$\vec{B}_{anis} = \frac{2K_{u1}}{M_s}(\vec{u}.\vec{m})\vec{u} \tag{4}$$

Here $K_{u1}$ indicates first order uniaxial anisotropy constant and $\vec{u}$ stands for a unit vector in the anisotropy direction. The temperature effect is calculated by:

$$\vec{B}_{thermal} = \vec{\eta}(step)\sqrt{\frac{2\alpha k_B T}{M_s \gamma \Delta V \Delta t}} \tag{5}$$

Where T is the temperature (K), $\Delta V$ is the cell volume, $k_B$ is the Boltzmann constant, $\Delta t$ is time step and $\vec{\eta}(step)$ is a random vector from a standard normal distribution which is independent (uncorrelated) for each of the three cartesian co-ordinates. Its value is changed after every time step.

The parameters listed in table 1 were used for all of the 0K simulation cases. However, $M_s$ and DMI were varied under a constant effective PMA energy to get an $M_s$ range and minimum DMI value for which skyrmions were formed as the lateral dimension was reduced from 100 nm to 20 nm.

Table 1: List of parameters

| Exchange stiffness | 6 pJ/m [30] |
|---|---|
| Thickness | 0.6 nm |
| Gilbert damping coefficient | 0.05 |
| Cell size | 0.5nm×0.5nm×0.6nm |

In our study, the cell size chosen remains well within the range of exchange length ($\sqrt{\frac{2A_{ex}}{\mu_0 M_s^2}}$) for saturation magnetization ranging from 1.0 MA/m-1.8 MA/m.

## Acknowledgement

M. M. R., D. B. and J. A. were supported by the National Science Foundation (NSF) Software and Hardware Foundations (SHF) Small Grant 1909030.

## Author contributions

M. M. R., D. B. and J. A. contributed to conceiving the idea and writing the paper. M. M. R. performed the simulations and all authors discussed and analyzed the results.

**Competing interests:** The author(s) declare no competing interests.